# Structural, electronic properties and stability of *MX*enes $Ti_2C$ and $Ti_3C_2$ functionalized by methoxy groups

*Andrey N. Enyashin,* Alexander L. Ivanovskii*

Institute of Solid State Chemistry, Ural Branch of the Russian Academy of Sciences, 620990 Ekaterinburg, Russia

ABSTRACT: The properties of *MX*enes, a new group of quasi-two-dimensional *d*-metal carbide or nitride nanomaterials derived by chemical exfoliation from the MAX phases, can be very sensitive to the presence of surface functional groups. Herein, the *MX*enes $Ti_2C$ and $Ti_3C_2$ functionalized by methoxy groups are considered by means of the density functional theory tight-binding method. Their structural, electronic properties, and relative stability are discussed in comparison with related and experimentally fabricated hydroxy derivatives of *MX*enes.

## 1. INTRODUCTION

The discovery of graphene – a single-atom thick molecular layer of carbon – has stimulated the search of its inorganic (non-carbon) analogues such as molecular single layers of $MoS_2$ and $NbSe_2$ chalcogenides, hexagonal BN etc.[1,2] The fabrication of all these two-dimensional (2D) nano-sized materials was successful in many ways due to a genuine stratiform structure of the compounds and low cohesion energy between the layers, which allow one to imply simply exfoliation of their bulks. While the observation of continuous single monolayers as rare events is possible due to the honoured method of micromechanical cleavage,[3] the mass production methods employ exfoliation via preliminary intercalation of the bulk (e.g., by alkali-metal atoms) with subsequent sinkage of intercalate in water solvent[4] or via direct physical intercalation by the molecules of an appropriate solvent.[5] Obviously, the aforementioned methods would fail in the production of 2D forms of non-layered compounds, which are characterized by a high cohesion energy between the atomic planes.

Nevertheless, certain 2D segments of non-layered compounds can be fabricated in a round route *via* chemical exfoliation of quasi-2D materials. Recently, so-called *MX*enes - 2D segments of cubic transition metal carbides and carbonitrides have been exfoliated from the MAX phases.[6] MAX phases are the closest relatives of carbide (carbonitride) compounds. Their lattice is a combination of atomic planes of *p*-elements covalently embedded between carbide (carbonitride) fragments. Despite strong bonding within the lattice, these planes can be selectively washed away, which opens new prospects of preparation and characterization of other 2D nano-sized materials. Nowadays, quite a broad group of *MX*enes ($Ti_2C$, $Ti_3C_2$, $Ta_4C_3$,TiNbC, $Ti_3CN_x$ and some others) was synthesized, and some physical and chemical properties of these materials were studied experimentally[6-9] or predicted theoretically.[6,10-14]

Moreover, already the first experiments[6,7] revealed that the surface of fabricated *MX*enes is terminated by various functional groups. For example, a 2D material with the nominal composition $Ti_3C_2$ can be produced by extraction of Al sheets from the 321 MAX phase $Ti_3AlC_2$ after treatment with HF solution, following the reaction $Ti_3AlC_2 + 3HF = AlF_3 + 3/2H_2 + Ti_3C_2$. However, the accompanying reactions $Ti_3C_2 + 2H_2O = Ti_3C_2(OH)_2 + H_2$ and $Ti_3C_2 + 2HF = Ti_3C_2F_2 + H_2$ simultaneously lead to the formation of hydroxylated or fluorinated $Ti_3C_2$ derivatives.[6]

The presence of surface functional groups may lead to drastic changes in the properties of *MX*enes. For example, unlike pristine $Ti_2C$ and $Ti_3C_2$ layers, which should act as magnetic metals, their F- or OH-functionalized derivatives can behave as non-magnetic narrow-band gap semiconductors or metals depending on the type of surface termination.[6,11-13] These results suggest the design of a rich variety of new 2D materials with different electronic, magnetic properties, and reactivity by modulating the type and degree of *MX*ene surface termination using various adatoms or molecules.

In this paper we conjecture the existence of a broad family of surface modified *MX*enes, which are closely related to the experimentally fabricated hydroxylated *MX*enes like $Ti_3C_2(OH)_2$ and could be also synthesized.[6,7] By analogy, the replacement of water solvent by an alcohol or alcohol water solutions could give rise to alkoxylated *MX*enes during acid treatment of MAX phases. For example, for $Ti_3AlC_2$, alkoxylated $Ti_3C_2$ derivatives might be obtained *via* direct reaction $Ti_3C_2 + 2ROH = Ti_3C_2(OR)_2 + H_2$ or *via* post-reaction of condensation $Ti_3C_2(OH)_2 + 2ROH = Ti_3C_2(OR)_2 + 2H_2O$, where R stands for an alkyl group. Thereinafter, we examine *MX*enes $Ti_2C$ and $Ti_3C_2$ functionalized by methoxy $OCH_3$ groups and focus on their structural, electronic properties, and relative stability, which are discussed in comparison with parent hydroxylated *MX*enes.

## 2. MODELS AND COMPUTATIONAL METHOD

The starting models of pristine *MX*enes $Ti_2C$ and $Ti_3C_2$ were constructed in a conventional way[11] by removing Al sheets from the parent MAX phases $Ti_2AlC$ and $Ti_3AlC_2$. These 2D materials consist of three ($Ti_2C$) and five ($Ti_3C_2$) atomic sheets with a hexagonal-like unit cell, where carbon sheets are sandwiched between the Ti sheets: [Ti-C-Ti] for $Ti_2C$ and [Ti2-C-Ti1-C-Ti2] for $Ti_3C_2$. The atoms are placed in the following positions: Ti: 4*f* (⅓, ⅔, *u*); and C: 2*a* (0, 0, 0) for $Ti_2C$, and Ti1: 2*a* (0, 0, 0,); Ti2: 4*f* (⅔, ⅓, *v*); and C(N): 4*f* (⅓, ⅔, *u*) for $Ti_3C_2$.

Next, the atomic models of the functionalized *MX*enes were designed assuming their complete surface termination by methoxy groups. Here, two main configurations of $OCH_3$ covering the external Ti sheet are admissible, when all methoxy groups are placed at the hollow sites between three neighboring carbon atoms within the carbide layer (position A), or these groups are placed at the top site of the carbon atoms (position B). Thus, for the double-sided surface termination of *MX*enes, three main configurations exist, namely, two symmetric configurations (AA and BB) and the asymmetric configuration AB, with compositions $Ti_2C(OCH_3)_2$ and $Ti_3C_2(OCH_3)_2$ see Figure 1. In all cases, the oxygen atom in the $OCH_3$ group has a tetrahedral coordination. Besides, we have also probed the location of the methoxy groups at the top site of the Ti atoms (position C), *i.e.* double coordination of oxygen atom in $OCH_3$ group, and some possible configurations for the double-sided surface termination: CC and AC, Figure 1. For comparison, we considered the related hydroxylated *MX*enes $Ti_2C(OH)_2$ and $Ti_3C_2(OH)_2$ and the system $Ti_2C(OCH_3)(OH)$, which simulates a more complex MXene derivative with a "mixed" type of covering: by $OCH_3$ and OH groups on the opposite surfaces.

All calculations of the structural and electronic properties of the modified MXene layers were performed within the density-functional based tight-binding method (DFTB) as implemented in the Dylax and deMon software.[15-17] This method was earlier applied for the simulations of various titanate nanostructures and organic adsorbates on titanium oxide.[18-20] It was found that these systems are described in reasonable agreement with experimental data and high-level theoretical methods at lower computational cost. All the structures considered in this work were fully relaxed under periodic boundary conditions as implemented in the Dylax software using the conjugate gradient algorithm and 10 *k*-points for each direction of two-dimensional reciprocal lattice. The self-consistent calculations were considered to be converged when the difference in the total energy did not exceed $10^{-3}$ eV/atom as calculated at consecutive steps and the forces between atoms are close to zero. Apart from the geometry optimization, possible reactivity of hydroxylated MXene $Ti_2C(OH)_2$ was tested using molecular-dynamics (MD) simulations within the framework of the same DFTB method in Γ-point approximation as implemented in the deMon software. The model chamber was represented by a periodic supercell composed of the optimized 5×5 supercell of $Ti_2C(OH)_2$ layer and the confined layer of methanol consisting of 36 $CH_3OH$ molecules (total thickness of these two layers was taken equal to 16 Å). This simulation box was annealed during 25 ps with the time step of 0.25 fs as an NVT-ensemble at temperature T = 300 K using a global Berendsen thermostat with the time constant s = 100 fs.

## 3. RESULTS AND DISCUSSION

**3.1. Structural properties.** As the first step, the full geometry optimization for free-standing $Ti_2C$ and $Ti_3C_2$ *MX*enes was carried out. The relaxed structures preserve their starting geometry and have the lattice parameters $a$ = 3.3338 and 3.2921 Å, respectively, in reasonable agreement with earlier estimations.[6,11] The lattice parameters of the hydroxylated forms $Ti_2C(OH)_2$ and $Ti_3C_2(OH)_2$ depend strongly on the type of OH covering and can be increased or decreased as compared with the parameters of $Ti_2C$ and $Ti_3C_2$, see the Table 1. If the surface OH groups are partially (in $Ti_2C(OCH_3)(OH)$) or fully replaced by methoxy groups (in $Ti_2C(OCH_3)_2$ and $Ti_3C_2(OCH_3)_2$), the lattice parameters increase as compared with the parameters of hydroxylated *MX*enes with the same type of surface covering. This trend is visualized in Figure 2, where the dependence of the lattice parameters for $Ti_2C$ and $Ti_3C_2$ functionalized by hydroxy and methoxy groups is depicted depending on the characteristic types of surface covering by these functional groups (AA, BB, and AB).

**3.2. Stability.** To provide an insight into the stability of the studied *MX*ene derivatives, we have estimated their model formation energies, $\Delta H$, assuming the formal reactions of methylation of free-standing $Ti_2C(OH)_2$ and $Ti_3C_2(OH)_2$. In this way, for example, for the reaction: $Ti_3C_2(OH)_2 + 2CH_3OH = Ti_3C_2(OCH_3)_2 + 2H_2O$, the value of $\Delta H$ was calculated as $\Delta H(Ti_3C_2(OCH_3)_2) = E_{tot}(Ti_3C_2(OCH_3)_2) + 2E_{tot}(H_2O) - E_{tot}(Ti_3C_2(OH)_2) - 2E_{tot}(CH_3OH)$, where $E_{tot}$ are the total energies of the corresponding substances as obtained in our DFTB calculations. Within this definition, a negative $\Delta H$ indicates that it is energetically favorable for given reagents to form more stable products, and *vice versa*. It is noteworthy that all of these calculations are performed at 0 K and zero pressure. The results listed in the Table reveal that all of the considered configurations of $Ti_2C(OCH_3)_2$ and $Ti_3C_2(OCH_3)_2$, excluding the most unfavorable configurations AC and CC, adopt negative values of $\Delta H$ suggesting the possibility of the fabrication of these materials leastwise by the condensation reaction. Note that not yet fabricated $Ti_2C$ derivatives with a “mixed” type of covering (by $OCH_3$ and OH groups on the opposite surfaces) also have a negative $\Delta H$ value indicating the possibility of such non-uniform functionalization of *MX*enes.

Additional information about the stability of different configurations of $Ti_2C$ and $Ti_3C_2$ *MX*enes depending on the type of their covering by methoxy groups can be received from their relative total energies, $\Delta E_{tot}$, listed in the Table. These data reveal that for both $Ti_2C(OCH_3)_2$ and $Ti_3C_2(OCH_3)_2$, the location of methoxy groups at the positions A, *i.e.* above the hollow sites between the three neighboring carbon atoms within the *MX*ene layer, is the most favorable one. On the contrary, the position C with a methoxy group above a Ti atom becomes most unstable. The position B, where the $(OCH_3)_2$ groups are placed at the top site of the carbon atoms, represents the intermediate case. As a result, for the case of double-side covering by methoxy groups, the stability of the *MX*ene methoxy derivatives decreases in the sequence: AA > AB > BB > AC > CC. Note that this stability hierarchy of the admissible configurations for $Ti_2C(OCH_3)_2$ and $Ti_3C_2(OCH_3)_2$ is fair also for hydroxylated derivatives of the discussed *MX*enes, see the Table 1.

**3.3. Electronic properties.** The electronic band structures, total and partial densities of states (DOSs) for the discussed *MX*ene derivatives are depicted in Figures 3 and 4. For the majority of these materials, the electronic bands are crossed by the Fermi level, $E_F$, indicating their metallic-like behavior, Figure 4. The exceptions are the compounds with lower stability: $Ti_2C(OCH_3)_2$ with configurations AB and AC, which can be characterized as narrow-gap semiconductors. Remarkably, in the cases of $Ti_3C_2(OCH_3)_2$ with configuration AB, and $Ti_2C(OH)(OCH_3)$ with configuration BB the valence and conduction bands meet at the Fermi level, Figure 4. This feature can point to a very small density of states at the Fermi level (like for semi-metals) and to the existence of Dirac points in these materials - like for graphene, see review [21].

In general, the valence spectra for all *MX*ene derivatives are quite similar. Therefore, we will discuss their peculiarities in more details on the example of $Ti_2C(OCH_3)_2$ with the most stable configuration AA. The valence states are divided into several subbands separated by the gaps, where the lowest-lying states are located below -8.5 eV and are formed mainly by the carbon and oxygen states of the methoxy groups, with admixtures of carbon and Ti states of $Ti_2C$ (Figure 5). The next subband comprises predominantly the oxygen states of the methoxy groups with additions of Ti states. These states correspond to the O-Ti bonds between methoxy groups and $Ti_2C$. The subband placed around -2.5 eV is of a hybridized (Ti *s,p,d* - C 2*p*) type, and these states are responsible for covalent Ti-C bonds within $Ti_2C$. Finally, the near-Fermi subband is composed mainly of the Ti 3*d* states.

The influence of the arrangement of methoxy groups on the electronic spectrum of *MX*ene derivatives can be illustrated by comparing the aforementioned DOSs picture with that of $Ti_2C(OCH_3)_2$ with the configuration CC, Figure 5. This type of configuration differs most essentially from the other types, since oxygen atoms have lower coordination numbers. The most pronounced differences concern the states of the atoms from the methoxy groups, which in this configuration are shifted by 1-2 eV higher the Fermi level and touch the bottom of the hybridized (Ti *s,p,d* - C 2*p*) band. In contrast, in the substitution of $Ti_3C_2$ for the *MX*ene lattice, i.e. comparing $Ti_2C(OCH_3)_2$ and $Ti_3C_2(OCH_3)_2$ with the same arrangement of methoxy groups AA, the positions of the valence bands remain almost unchanged, and the main differences can be attributed to the difference in the states in pure $Ti_2C$ and $Ti_3C_2$ *MX*enes (for more details about these differences for pristine *MX*enes see Ref. [11]).

**3.4. Interactions in $Ti_2C(OH)_2$ - $CH_3OH$ system.** Taking into account the metallic-like character of the hydroxylated *MX*enes, one may expect their high reactivity, namely, the ability to exchange ions (protons or OH groups). Meanwhile, alcohols demonstrate either basic or acidic properties. In this section, we discuss and give a preliminary look at the possible mechanism of functionalization of *MX*enes by alkoxy groups after the condensation reaction between hydroxylated *MX*enes and alcohols. For this purpose, MD simulation of the interaction between hydroxylated *MX*ene $Ti_2C(OH)_2$ and methanol, $CH_3OH$, was carried out as described in Computational part.

Despite some naivety of this quantum-chemical MD simulation, a couple of important effects can be distinguished at the interface of hydoxylated *MX*enes and alcohols. A detailed visualization of the equilibrated system uncovers the products of two main reactions in the model chamber (Figure 6). The appearance of protonated $CH_3OH_2^+$ molecules can be observed due to the transfer of protons from the hydroxy groups of $Ti_2C(OH)_2$ surface. Simultaneously, a few $CH_3OH$ molecules form surface complexes with Ti atoms of $Ti_2C(OH)_2$ via anchoring by hydroxyl groups. We are not aware of the most appropriate reaction conditions (temperature, pressure) and are limited by accessible simulation time and the size of molecule ensemble.

However, already this demonstrates a high reactivity of hydroxylated *MX*enes. It is possible to surmise that these products can be intermediate and a set of consecutive reactions can be suggested.

First, protonated $CH_3OH_2^+$ molecules can attack surface hydroxy groups with the formation of surface methoxy groups and liberation of water molecules and protons. This mechanism could lead to the formation of suggested methoxylated *MX*ene like $Ti_2C(OCH_3)_2$ following the formal equation $Ti_2C(OH)_2 + 2CH_3OH = Ti_2C(OCH_3)_2 + 2H_2O$. Besides, a competing process of dimethyl ether $CH_3OCH_3$ formation may be expected due to the reaction between electrophilic $CH_3OH_2^+$ and neutral $CH_3OH$ molecules, which points to possible catalytic activity of hydroxylated *MX*enes in respect to the formal equation $2CH_3OH = CH_3OCH_3 + H_2O$.

Second, a $CH_3OH$ molecule adsorbed on Ti atoms could react with one of three hydroxy groups anchored to the same Ti atom. Then, the formation of $CH_3O$ group anchored to the Ti atom and liberation of water molecule can be expected with a final product of functionalized *MX*ene $Ti_2C(OCH_3)_2$.

## 4. CONCLUSIONS

In conclusion, we have examined the structural and electronic properties and comparative stability of *MX*enes $Ti_2C$ and $Ti_3C_2$ functionalized by methoxy groups depending on the type of pristine *MX*enes: $Ti_2C$ *versus* $Ti_3C_2$ and the types of their termination by $OCH_3$ groups. Our calculations show that the relaxed $Ti_2C(OCH_3)_2$ and $Ti_3C_2(OCH_3)_2$ structures preserve their 2D-like geometry testifying to their stability, and their lattice parameters depend mainly on the type of $OCH_3$ covering.

The data obtained reveal the stability hierarchy of the accessible configurations for $Ti_2C(OCH_3)_2$ and $Ti_3C_2(OCH_3)_2$. Namely, for both *MX*ene derivatives, the location of methoxy groups above the hollow sites between the three neighboring carbon atoms and tetrahedral coordination of O atoms is the most favorable one, while the location of these groups directly above the Ti atoms and the double coordination of O atoms is least stable. *MX*ene derivatives with a “mixed” type of covering by $OCH_3$ and OH groups demonstrate a negative formation energy indicating a possibility of such non-uniform functionalization of *MX*enes.

The most stable *MX*ene derivatives are metallic-like materials, in which the near-Fermi electronic bands are composed mainly of Ti 3*d* states. On the other hand, for some meta-stable configurations, the cross points of valence and conduction bands found at the Fermi level suggest the existence of Dirac points for these materials - like for graphene.

Finally, the formation of the supposed materials was probed by means of MD simulations on the example of interface between the hydroxylated MXene layer $Ti_2C(OH)_2$ and methanol. Possible binding of $CH_3OH$ molecules with the surface Ti atoms of *MX*ene by their OH groups was confirmed. Besides, these simulations pointed to high protonating activity of hydroxylated *MX*enes, which allows us to consider these compounds as perspective catalysts, in particular, for esterification reactions. In turn, we can speculate that alkoxylated *MX*enes may be promising compounds as the materials of Li-ion batteries with a higher Li-ion mobility and a higher affinity to organic electrolytes due to their hybrid organic-inorganic character. As well, these compounds could be considered as the materials for catalysis of enantioselective reactions like their molecular counterparts.[22] Future experiments on the fabrication of the suggested alkoxi *MX*enes and study of their properties as well as massive molecular dynamics simulations in the framework of a more simple scheme like reactive force-field methods with careful parametrization[23] could prove our suggestions.

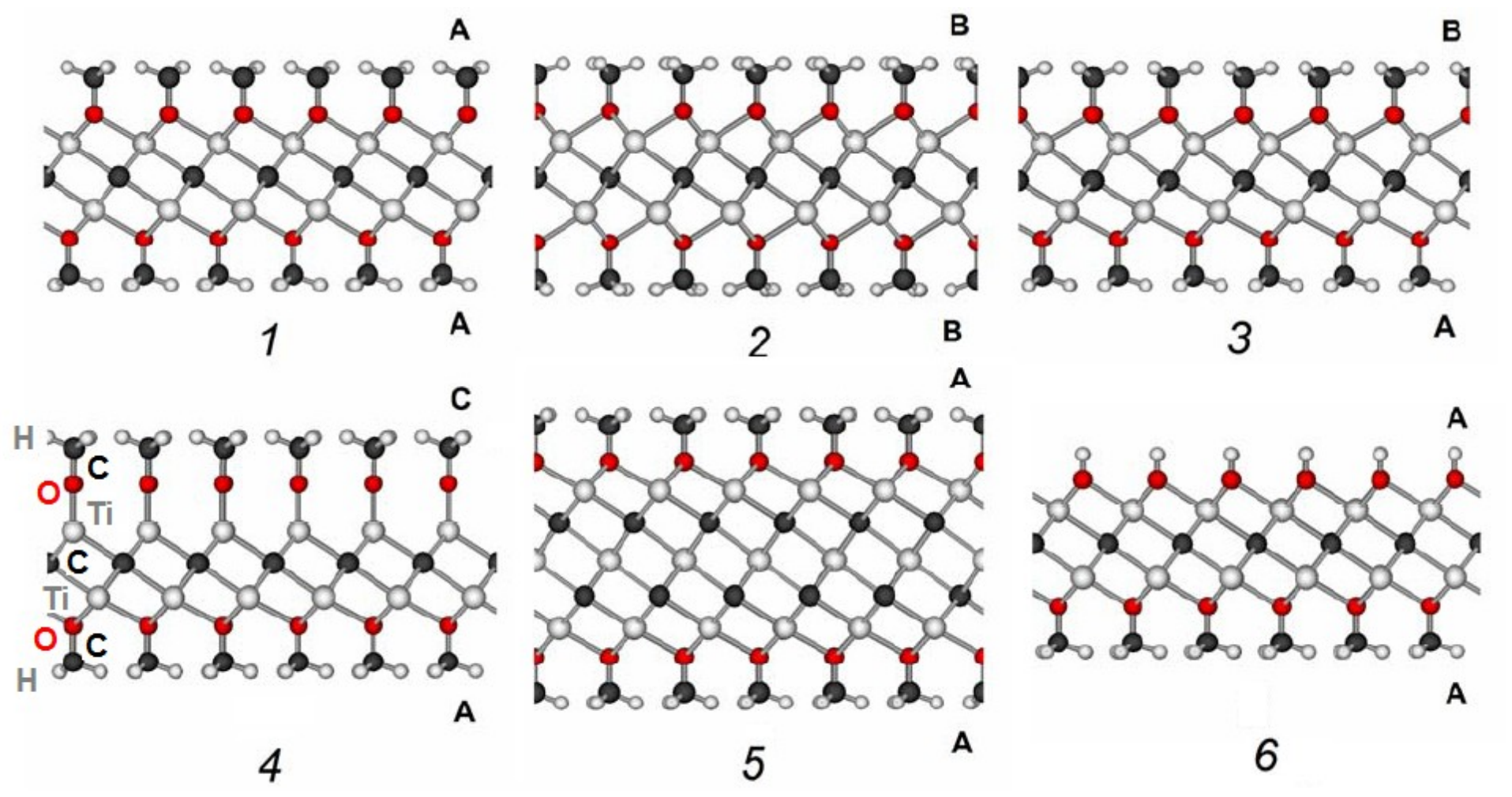

**Figure 1.** Optimized atomic structures of examined *MX*enes $Ti_2C$ and $Ti_3C_2$ functionalized by methoxy groups: $Ti_2C(OCH_3)_2$ (*1*-*4*), $Ti_3C_2(OCH_3)_2$ (*5*) and a $Ti_2C$ derivative with a “mixed” type of covering (by $OCH_3$ and OH groups on the opposite surfaces): $Ti_2C(OH)(OCH_3)$. The considered configurations of $OCH_3$ (OH) covering: A - all methoxy (hydroxy) groups are placed at the hollow site between three neighboring carbon atoms $C_3$ within *MX*ene layer; B - these groups are placed at the top site of the carbon atom; and C - these groups are placed at the top site of the Ti atom.

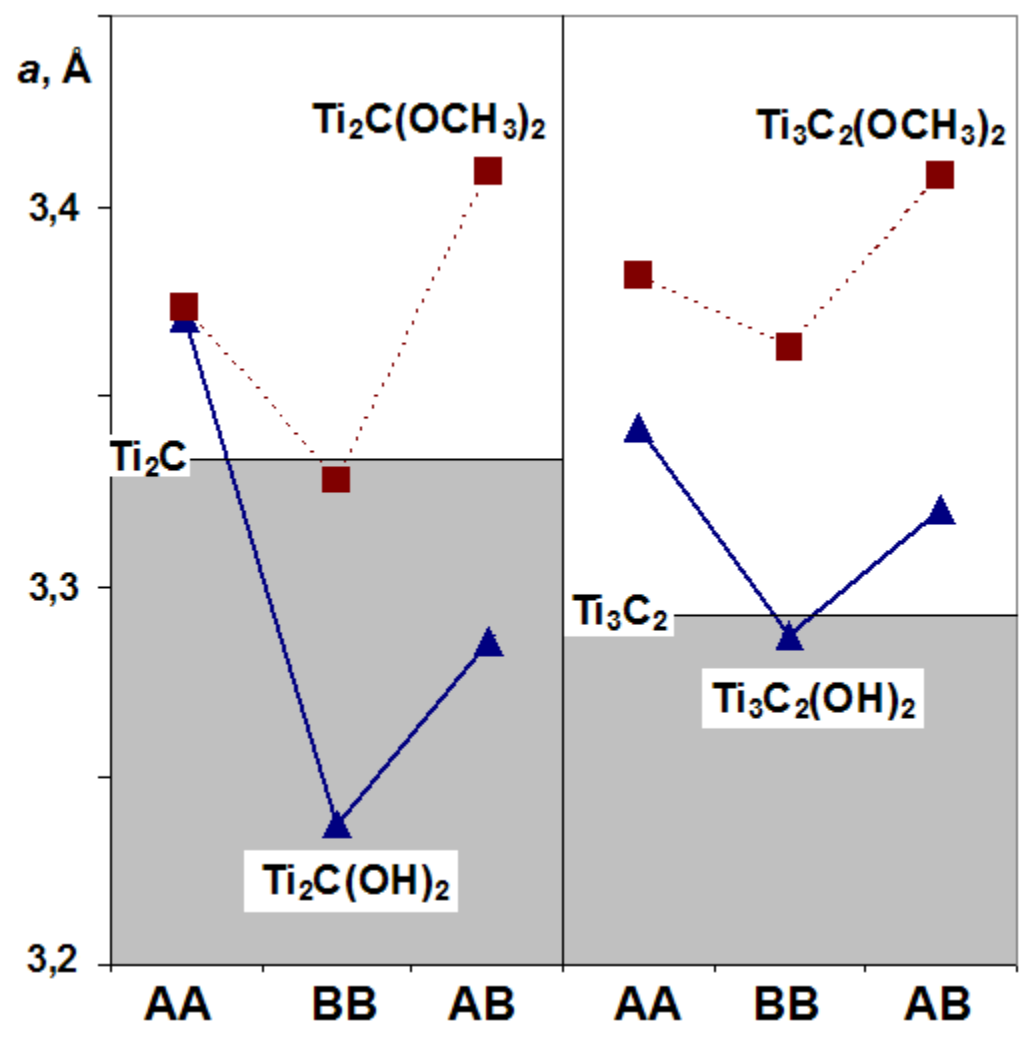

**Figure 2.** Optimized lattice parameters for $Ti_2C$ and $Ti_3C_2$ functionalized by hydroxy and methoxy groups as depending on the types of surface covering by these functional groups (symmetric configurations AA and BB and asymmetric configuration AB, see Fig. 1) in comparison with pristine $Ti_2C$ and $Ti_3C_2$.

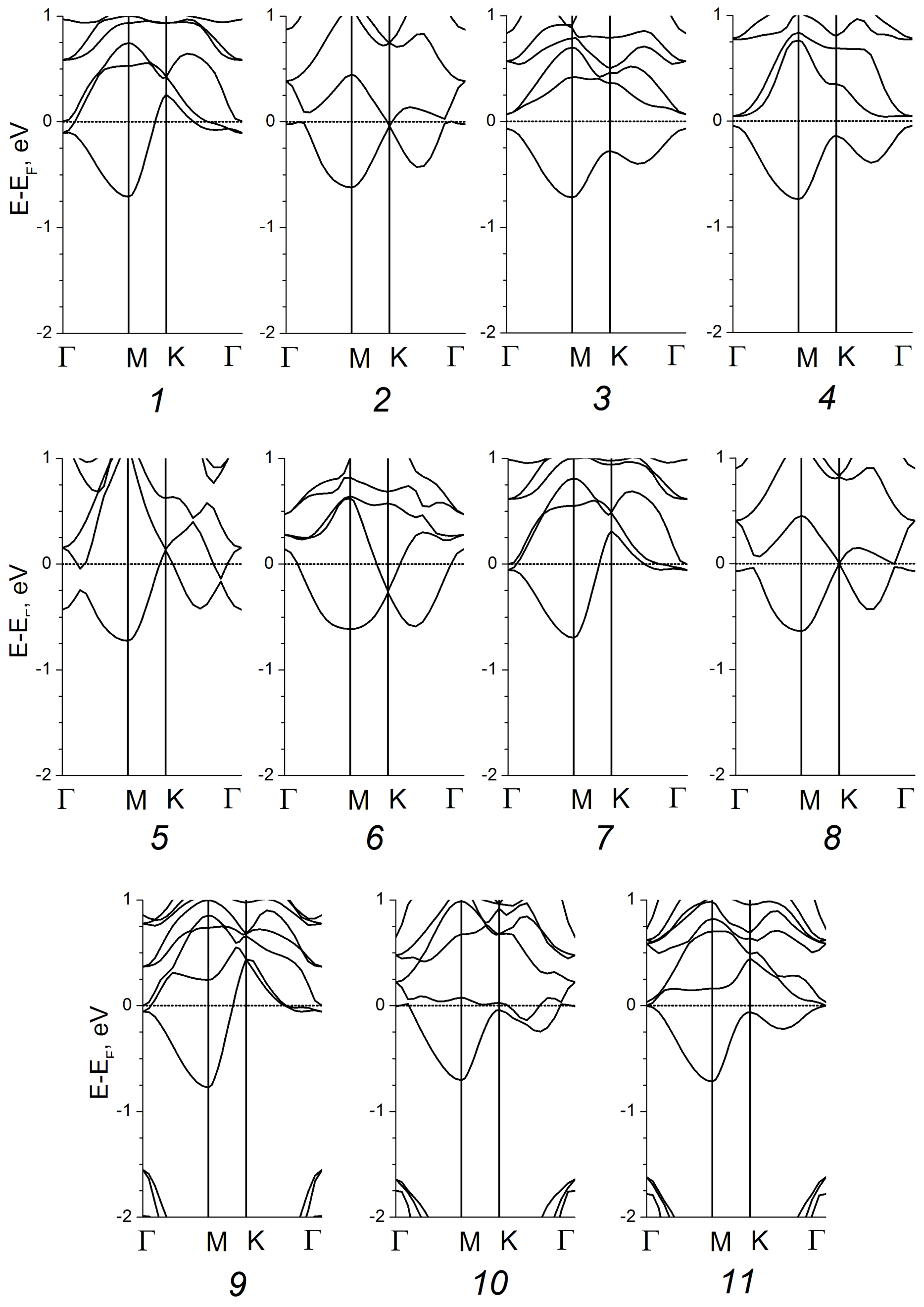

**Figure 3.** Near-Fermi electronic bands for methoxylated $Ti_2C$ (*1*-*6*) and $Ti_3C_2$ (*9*-*11*) with different types of surface covering: AA (*1*), BB (*2*), AB (*3*), AC (*4*), CC (*5*), BC (*6*), AA (*9*), BB (*10*), and AB (*11*). The electronic bands for $Ti_2C$ derivative with a "mixed" type of covering of $Ti_2C(OH)(OCH_3)$ with configurations AA (*7*) and BB (*8*) are also depicted.

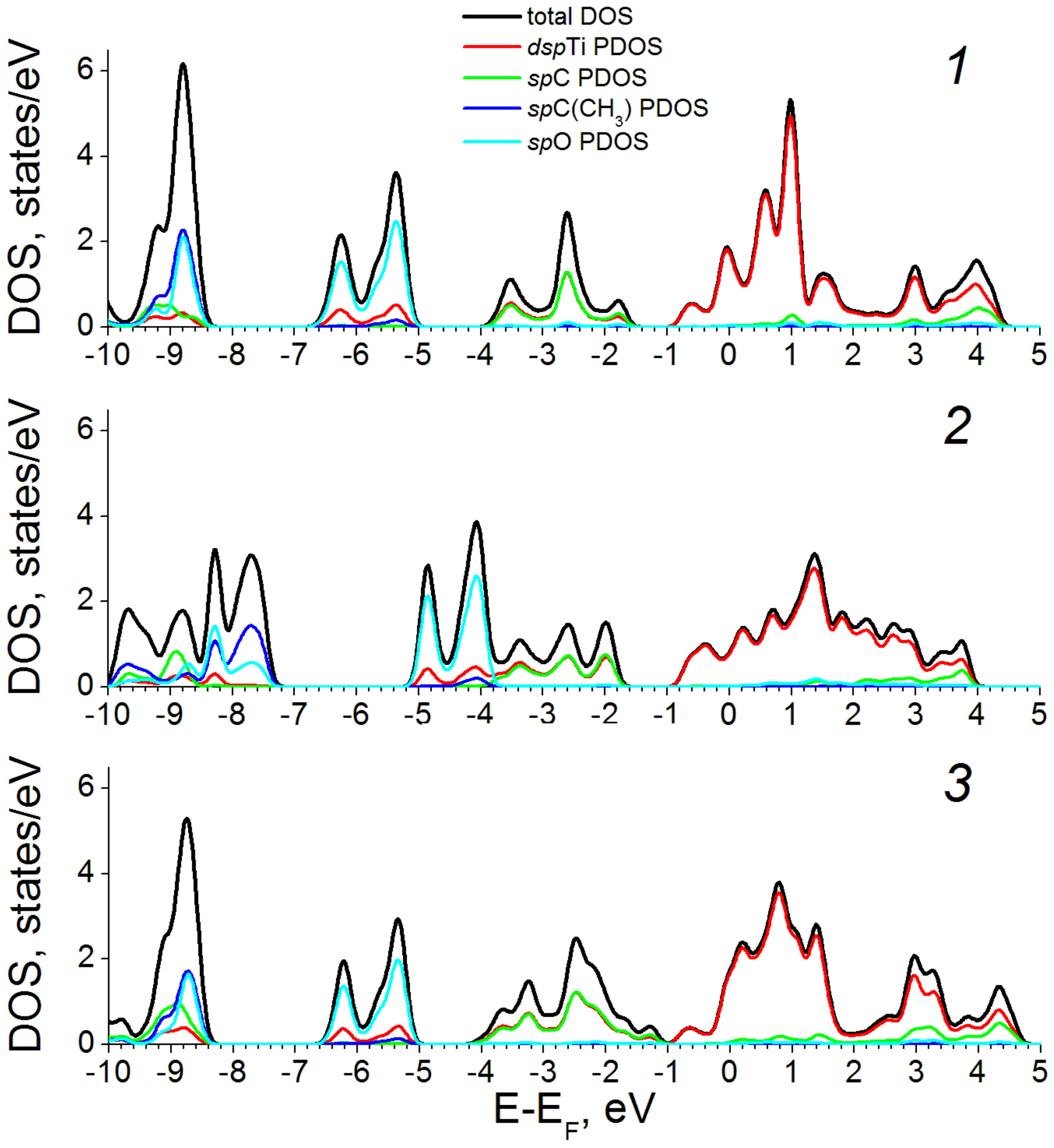

**Figure 4.** Total and atomic-resolved densities of states for $Ti_2C(OCH_3)_2$ with configuration of surface groups AA (*1*), $Ti_2C(OCH_3)_2$ with configuration CC (*2*), and $Ti_3C_2(OCH_3)_2$ with configuration AA (*3*).

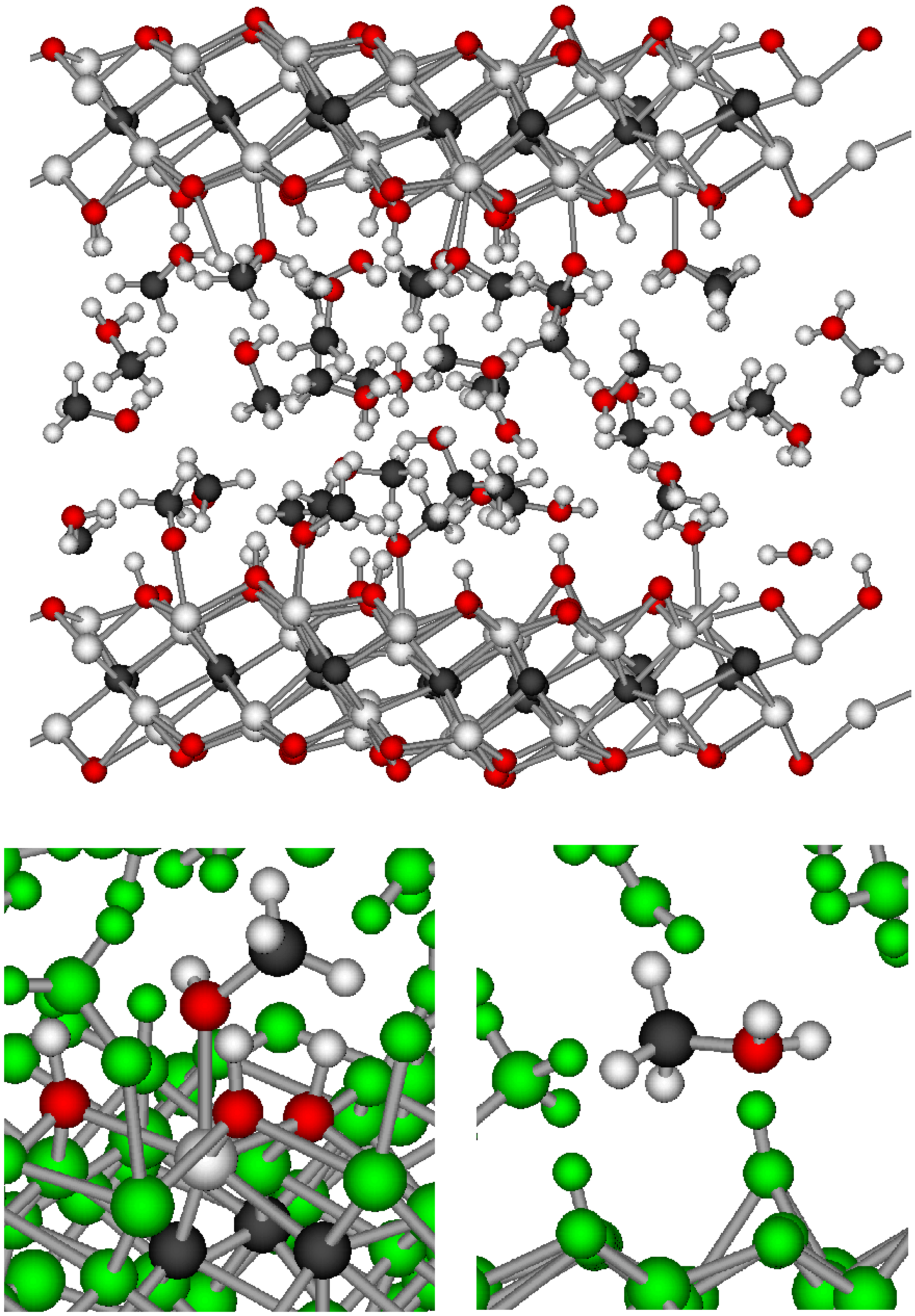

**Figure 5.** Screenshot of the DFTB molecular dynamics simulation for the system "$Ti_2C(OH)_2$ – methanol" at 300 K after 25 ps (on the top). A more detailed vizualization demonstrates the occurrence of protonated $CH_3OH$ molecules (bottom right) and $CH_3OH$ molecules anchored by O atom to the Ti surface of $Ti_2C(OH)_2$ layer (bottom left).

**Table 1.** Optimized lattice parameters (*a*, in Å), relative energies (Δ*E*, eV/atom), and model formation energies (Δ*H*, eV/$OCH_3$) for $Ti_2C$ and $Ti_3C_2$ functionalized by methoxy and hydroxy groups as obtained within DFTB calculations.

| system * | *a* | Δ*E* | Δ*H* |
|---|---|---|---|
| $Ti_2C(OCH_3)_2$ (AA) | 3.373 | 0 | -0.335 |
| $Ti_2C(OCH_3)_2$ (BB) | 3.328 | 0.042 | -0.240 |
| $Ti_2C(OCH_3)_2$ (AB) | 3.409 | 0.034 | -0.234 |
| $Ti_2C(OCH_3)_2$ (AC) | 3.385 | 0.106 | 0.355 |
| $Ti_2C(OCH_3)_2$ (CC) | 3.390 | 0.215 | 1.060 |
| $Ti_3C_2(OCH_3)_2$ (AA) | 3.382 | 0 | -0.357 |
| $Ti_3C_2(OCH_3)_2$ (BB) | 3.363 | 0.044 | -0.283 |
| $Ti_3C_2(OCH_3)_2$ (AB) | 3.408 | 0.031 | -0.282 |
| $Ti_2C(OCH_3)(OH)$ (AA) | 3.363 | 0 | -0.331 |
| $Ti_2C(OCH_3)(OH)$ (BB) | 3.296 | 0.045 | -0.241 |
| $Ti_2C(OH)_2$ (AA) | 3.371 | 0 | - |
| $Ti_2C(OH)_2$ (BB) | 3.237 | 0.054 | - |
| $Ti_2C(OH)_2$ (AB) | 3.285 | 0.020 | - |
| $Ti_3C_2(OH)_2$ (AA) | 3.342 | 0 | - |
| $Ti_3C_2(OH)_2$ (BB) | 3.287 | 0.072 | - |
| $Ti_3C_2(OH)_2$ (AB) | 3.320 | 0.030 | - |

* the types of configurations (AA, AB *etc*) see in Fig. 1.

AUTHOR INFORMATION

**Corresponding Author**

*E-mail: enyashin@ihim.uran.ru

ACKNOWLEDGMENTS

We acknowledge financial support from RFBR (the grant 11-03-00156-a).